\documentclass[a4paper,unpublished]{quantumarticle}
\pdfoutput=1
\usepackage[numbers,sort&compress]{natbib} 
\usepackage{graphicx} 
\usepackage[svgnames,dvipsnames]{xcolor}
\usepackage{braket}
\usepackage{soul}
\usepackage{amsmath}
\usepackage{dcolumn}
\usepackage{bm}
\usepackage{hyperref}
    \hypersetup
    {   
        unicode=true,           
        pdftoolbar=true,        
        pdfmenubar=true,        
        pdffitwindow=false,     
        pdfstartview={FitH},    
        pdfauthor={Edoardo Zavatti},
        pdftitle={More is uncorrelated: Tuning the local correlations of SU(N) Fermi-Hubbard systems via controlled symmetry breaking},
        pdfkeywords={metal-insulator transition, entanglement measures, correlation measures, nonfreeness, nongaussianity, quantum information theory, dynamical mean-field theory, strong correlations, multi-component systems, cold atom simulators, SU(N)},
        pdfnewwindow=true,      
        colorlinks=true,        
        linkcolor=Maroon,       
        citecolor=Cerulean,     
        filecolor=Maroon,       
        urlcolor=Cerulean       
    }
\usepackage[capitalize]{cleveref}
\crefname{equation}{Eq.\!\!}{Eqs.\!\!}
\crefname{figure}{Fig.\!}{Figs.\!}
\crefname{section}{Sec.\!}{Secs.\!}
\usepackage{tikz}
\usepackage{mathtools}
\usepackage{gensymb}
\usepackage{amssymb}
\usepackage{xfrac}
\usepackage{siunitx}
\usepackage[utf8]{inputenc}
\usepackage[autostyle]{csquotes}

\usepackage{ragged2e}
\makeatletter
\renewcommand{\@makecaption}[2]{%
  \vskip\abovecaptionskip
  \justifying
  \small #1: #2\par
  \vskip\belowcaptionskip}
\makeatother

\usepackage[normalem]{ulem} 

\begin{document}

\title{More is uncorrelated: Tuning the local correlations of SU($N$) Fermi-Hubbard systems via controlled symmetry breaking} 

\author{E. Zavatti}
\email{ezavatti@sissa.it}
\affiliation{Scuola Internazionale Superiore di Studi Avanzati, Trieste, Italy}
\author{G. Bellomia}
\email{gabriele.bellomia@tuwien.ac.at}
\affiliation{Scuola Internazionale Superiore di Studi Avanzati, Trieste, Italy}
\affiliation{%
    Institute of Solid State Physics, TU Wien, Vienna, Austria}
\author{M. Ferraretto}
\affiliation{Scuola Internazionale Superiore di Studi Avanzati, Trieste, Italy}
\author{S. Giuli}
\affiliation{Scuola Internazionale Superiore di Studi Avanzati, Trieste, Italy}
\affiliation{Center for Computational Quantum Physics, Flatiron Institute, New York, USA}
\author{M. Capone}
\affiliation{Scuola Internazionale Superiore di Studi Avanzati, Trieste, Italy}
\affiliation{Istituto Officina dei Materiali, Consiglio Nazionale delle Ricerche, Trieste, Italy}


\begin{abstract}

Cold-atom experiments based on alkali-like atoms provide us with a tool to experimentally realize Hubbard models with a large number $N$ of components. The value of $N$ can be seen as a new handle to tune the properties of the system, leading to new physics both in the case of fully SU($N$) symmetric systems, or in the presence of controlled symmetry breaking.
We focus on the Mott transition at global half filling and we characterize local correlations between particles through the \emph{inter-flavor mutual information}, an experimentally accessible quantity that rigorously measures the distance from the closest gaussian state, unveiling features that cannot be accessed by conventional probes of Mottness.
We prove that these correlations are fully independent from local entanglement and quantum discord, and,
using Dynamical Mean-Field Theory, we show that the SU(4) system has significantly smaller correlations than the SU(2) counterpart. In the atomic limit we prove that increasing $N$ further decreases the strength of the correlations.
This suggests that a controlled reduction of the symmetry, reducing the number of effective components, can be used to enhance the degree of correlation.
We confirm this scenario solving the model for $N=4$ and gradually breaking the symmetry via a Raman field, revealing an evolution from the SU(4) to the SU(2) Mott transition as the symmetry-breaking term increases, with a sudden recovery of the large correlations of the SU(2) model at weak Raman coupling in the Mott state.  By further exploring the interplay between energy repulsion and the Raman field, we obtain a rich phase diagram with three different phases -- a metal, a band insulator, and a Mott insulator -- all coexisting at a single \emph{tricritical point}.
\end{abstract}

\maketitle

\section{\label{sec: introduction} Introduction}
It is hard to overstate the importance of the fermionic Hubbard model \cite{Hubbard1963}, a deceivingly simple model originally proposed to understand itinerant magnetism, that became perhaps the most studied model in condensed matter when its two-dimensional version was proposed to explain high-temperature superconductivity. 

Even if the fight to solve the standard spin-1/2 Hubbard model with SU(2) symmetry in two dimensions is still underway, generalizations of the model including more degrees of freedom have shown to display new fundamental physics both in a solid-state framework \cite{georges_hundcoupling} and in quantum simulators. Some notable examples are interaction-resilient metals \cite{Janus2011,Isidori2019,Richaud2021,Richaud2022} and selective Mott states, in which some flavors are Mott localized and others are metallic \cite{LdM2005_orb-selective,LdM2014_orb-selective_key-iron,del2018selective,flavor_selective_experiment}.
In this context, ultracold atoms in optical lattices have flourished as powerful and flexible experimental platforms to simulate the multi-component Hubbard model, particularly leveraging the SU($N$) symmetry (associated to the nuclear spin) of alkaline-earth-like atoms such as $^{173}$Yb and $^{87}$Sr \cite{ultracold_SU(N),alkaline_SU(N),SUN_experiment_2,SUN_experiment_3,scazza_SU(N),SUN_experiment_4,Scazza_SU(N)_crossover,SUN_experiment_5,SUN_experiment_6,Bloch_SU(N)}. 

These systems also offer the possibility to  drive the system from the full SU($N$) symmetry to a lower one in a continuous and controlled way. This controlled symmetry breaking has allowed to explore flavor-selective  Mott localization, starting from an SU(3) model, from both the theoretical and experimental points of view
\cite{del2018selective,flavor_selective_experiment}.
Flavor-selective Mott localization is indeed the extreme version of flavor-selective correlations, i.e.\,the simultaneous presence of particles with different degree of correlation in the same quantum system. In this perspective, it can be seen as one example of the possibility to control the degree of correlation of quantum many-body systems via the controlled breaking of the SU($N$) symmetry.

In this work we build on these ideas and we put these concepts on more solid grounds assessing 
the strength and the nature of the correlation properties as the number of components $N$ varies.  

A wide palette of quantum-information related quantities have indeed been introduced to \emph{quantify} and \emph{classify} correlations in quantum many-body systems. Together with a conventional use to pinpoint and characterize the phase diagram of quantum many-body systems
\cite{amico2008entanglement,dunleavy2015mutual,lepori2022mutual,li1990mutual,calabrese2004entanglement,vidal2003entanglement,gokmen2021statistical,Hauke2016,Costa&Hauke_multipartite_coldatoms,GMazza_EPR_competing_orders,Held_QFI_pseudodap,entropy_MottMIT,walsh2019local,bellomia2024quasilocal,bellomia2024quantum,Roshap2025_timescale_entanglement}, these quantities have recently
gathered a renewed interest for the assessment of genuine resources for quantum advantage \cite{QResources_RMP}. 
Among the different proposals, \emph{entanglement} \cite{HHHH_RMP,Plenio_EntanglementMeasures,EntanglementDetection_review,nielsen2010quantum,Schilling_QuantumScience}, \emph{quantum discord} \cite{Discord_def,Vedral2001,Modi2010,Discord_state-prep,Discord_q-comp,Discord_q-crypto,Discord_q-thermodyn} and \emph{nongaussianity} \cite{Gottlieb2005,Gottlieb2007,Paris_nonGaussianity,GeometricEntropy_MIT_Vollhardt,Held_nonfreeness,LyuBu24_nonGaussianity,Sierant2025_FAF,Sierant2026_FAF,Matchgate_Review,Mele_NonfreenessBounds,Mele_NonfreenessLearning,Lami2025_nongauss_doping} are promising candidates to properly define the magnitude and the nature of correlations.

This program exploits the development of a unified geometrical approach \cite{Vedral_RMP} for the quantification and classification of resources in open quantum systems, that allows to evaluate the \emph{informational} distance of a given density matrix from reference gaussian states \cite{GeometricEntropy_MIT_Vollhardt,Held_nonfreeness,Paris_nonGaussianity}, separable states
\cite{EntanglementDetection_review,Plenio_EntanglementMeasures} or pseudo-classical states \cite{Vedral2001,Modi2010,Discord_def}. 

In a recent work it has indeed been shown that one can compute the local nonfreeness  (i.e.\,the correlation between fermions on the same site) as the mutual information between local natural spin orbitals \cite{Bellomia_intracorr}, a simple and experimentally accessible quantity. 


Here we leverage on this result to show that for multi-component fermionic models that conserve the number of particles in each flavor, such as the SU($N$) Hubbard model, the local correlations are effectively classical and they are not related to local entanglement, nor quantum discord. Hence, these correlations are associated, at the local level, with pure nongaussian resources.

By studying the  half-filled SU($N$) Hubbard model  (i.e.\,with $N/2$ fermions per site) by means of Dynamical Mean-Field Theory (DMFT)\,\cite{georges1996dynamical}, we show that these classical correlations are dramatically reduced going from the popular SU(2) model to SU(4). Solving exactly the atomic limit, we provide an evidence that this trend is more general, with the strength of the correlations decreasing with the number of components $N$, leading to an effective decoupling in the large-$N$ limit. 

We finally show that a controlled breaking of the SU(4) symmetry provides a tool to effectively reduce the number of components, hence increasing the correlations. This scenario can be experimentally realized and detected in cold-atoms based simulators of SU($N$) fermions.




%


After introducing the model in \cref{sec: model}, and the entropy-related quantities that we use to quantify correlations in \cref{sec: entropic_measures}, we consider the SU(4)-symmetric Hubbard model, comparing it to the SU(2) case, in \cref{sec: su2vssu4}. By means of inter-flavor mutual information, we show that correlations in the Mott phase are significantly weaker for the four-component system than for the two-component one. We then extend this result analytically in \cref{sec: sun}, by studying the atomic limit of an SU($N$)-symmetric Hubbard model, demonstrating that mutual information vanishes in the large-$N$ limit, leading to an uncorrelated Mott insulator. Finally, in \cref{sec: omega} we explore the effects of symmetry breaking, introducing a term that lifts the degeneracy between two of the four flavors. This induces a crossover from SU(4) to SU(2), tunable via a symmetry-breaking term.

\section{\label{sec: model} Model}
\begin{figure}
    \centering
    \includegraphics[width=1.0\linewidth]{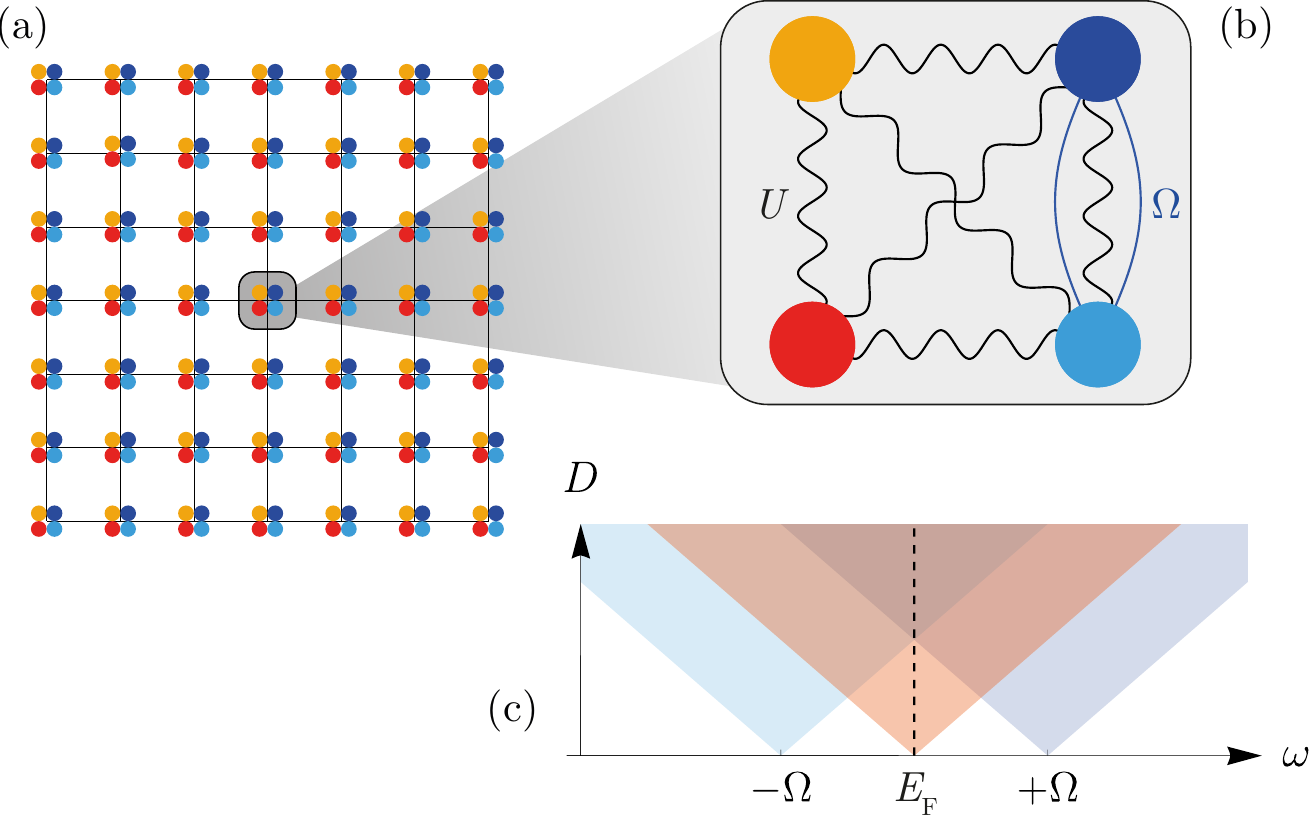}
    \caption{(a) Schematic representation of an SU(4) lattice, with nearest neighbors hopping. Each site can host up to 4 flavors. (b) The inter-flavor energy repulsion, equal for each pair of flavors, is labeled by $U$, while $\Omega$ represents a flavor-selective Raman field, which acts essentially as an inter-flavor hopping.
    (c) The evolution of the flavored dispersions as a function of the Bethe half-bandwidth 
    $D$, where the dashed line marks the Fermi level $E_\mathrm{F}$ and 
    $\pm\Omega$ marks the Raman-induced split of the coupled flavors.}
    \label{fig:system}
\end{figure}

Alkaline-earth-like fermionic atoms, such as $^{173}$Yb and $^{87}$Sr are an ideal platform to realize multicomponent Hubbard models, owing to the perfect decoupling between electronic and nuclear degrees of freedom due to the vanishing electronic angular momentum. As are result the scattering lengths in the  electronic ground state $^1S_0$~\cite{alkaline_SU(N),ferraretto_tesi} do not depend on the projection of the nuclear angular momentum. Hence, for and angular momentum $I$ ($I=5/2$ for $^{173}$Yb and $I=9/2$ for $^{87}$Sr), they have  a SU($2I+1$) symmetry.

Here we consider a balanced mixture with $N=4$ at global half-filling ($N/2 =2$ fermions per lattice site) out of the $2I+1$ available nuclear states loaded into an optical lattice, whose depth controls the ratio $U/t$, thus realising a SU(4)-symmetric tunable Hubbard model. The Hamiltonian of the model reads
\begin{multline}    \label{eq: H}
    \mathcal{H}=-t\sum_{\langle i,j\rangle, \alpha}\left({c^\dagger_{i,\alpha}c_{j,\alpha}+\text{h.c.}}\right)+\frac{U}{2}\sum_{i,\alpha\neq \beta}n_{i,\alpha}n_{i,\beta}, 
\end{multline}
where $c_{i,\alpha}$ ($c^\dagger_{i,\alpha}$) destroys (creates) a fermion at site $i$ and with flavor $\alpha \in \{1, 2, 3, 4\}$, and $n_{i,\alpha}=c^{\dagger}_{i,\alpha}c_{i,\alpha}$ is the corresponding number operator.

The SU(4) symmetry of the interaction can be broken explicitly via laser-induced two-photon Raman processes that stimulate the electronic transition $^1S_0\leftrightarrow$ $^3P_1$. This leads to a tunneling between different flavors, which can be tuned and made flavor-selective~\cite{flavor_selective_experiment,chiral_currents_Science, Mancini_thesis,spinorbit_theory}. In particular we couple only two flavors  according to
\begin{equation}
\label{H_Raman}
    H_\mathrm{Raman} = \Omega \sum_i \left( c^\dagger_{i,3}c_{i,4} + c^\dagger_{i,4}c_{i,3}\right),
\end{equation}  
where the amplitude $\Omega$ can be tuned by varying the intensity of the two laser beams.
The model is schematically shown in \cref{fig:system}. We refer to flavors $|3\rangle$ and $|4\rangle$ as the \emph{Raman-coupled} flavors, and flavors $|1\rangle$ and $|2\rangle$ as the \emph{Raman-free} flavors.

We solve the model at zero temperature using DMFT. 
Within DMFT, the lattice model is mapped onto an effective impurity model, which must be solved self-consistently by ensuring that the impurity Green’s function matches the local component of the lattice Green’s function. As often done, we consider a Bethe lattice with infinite coordination, whose density of states $N(\epsilon)=\frac{2}{\pi D^2}\sqrt{D^2-\epsilon^2}$ has a compact support despite the infinite-coordination limit. 
The impurity model is solved using the Lanczos/Arnoldi exact diagonalization (ED) method \cite{caffarel1994exact,capone2007solving,amaricci2022edipack,EDIpack2,EDIpack_code}.  
Here we use $N_\text{bath}=3$ bath sites (each with $N=4$ flavors), which allows for reasonably cheap calculations.  We also verified that using a larger number of bath sites did not lead to substantial modifications in the results, as documented in previous works \cite{del2018selective,georges1996dynamical}.

For $\Omega=0$, as mentioned above, we recover the  SU(4) Hubbard model which, as previously studied within DMFT, 
undergoes a metal-to-insulator transition (the Mott transition) when $U\simeq 4D$, where $D\propto t$ is the half-bandwidth~\cite{Koga2002,Koga2015,Unukovych2021,Yanatori2016,giuli2023mott,Golubeva2017,Lee2018}.  
For finite $\Omega$, Eq.\,(\ref{H_Raman}) can be diagonalized straightforwardly site by site in terms of bonding and antibonding combinations of the Raman-coupled species. Denoting the new basis as $\{ \tilde{c}_{i,\alpha}, \tilde{c}^{\dagger}_{i,\alpha} \}$ the Raman coupling becomes an energy splitting (or flavor-selective magnetic field), as shown in panel (c) of \cref{fig:system}:
\begin{equation}
    \Omega \sum_i \left( \tilde{n}_{i,3} - \tilde{n}_{i,4} \right), 
\end{equation}
while the Hubbard interaction is invariant under rotations. In the rest of this work we will use this rotated basis both for physical transparency and for practical reasons. The local single-fermion correlation matrix is indeed diagonal in this basis, so that the corresponding single-fermion orbitals are the local \emph{natural} orbitals of the system \cite{Natural_spin-orbitals_def,Faraday_Natural_Orbitals}. 

At $U=0$, the Raman-coupled flavors form bands centered at $\pm\Omega$, while Raman-free flavors form two degenerate bands centered at the Fermi level.

As $\Omega$ increases, the Raman-coupled flavors become, respectively, full and empty. 
Hence they are effectively decoupled from the Raman-free levels centered at zero energy, which realize an effective SU(2) model. Consequently, the $\Omega$-term induces an evolution from the SU(4)-symmetric Hubbard model ($\Omega=0$) to an effective SU(2)-symmetric model, when $\Omega\gtrsim D$. 


\section{\label{sec: entropic_measures} Assessing inter-flavor correlations}

The von Neumann mutual information is a fundamental measure of
both classical and quantum correlations between two partitions of a quantum system \cite{HHHH_RMP,Plenio_EntanglementMeasures,QResources_RMP,EntanglementDetection_review,nielsen2010quantum,Schilling_QuantumScience},
which has been already widely used in different frameworks \cite{amico2008entanglement,dunleavy2015mutual,lepori2022mutual,li1990mutual,calabrese2004entanglement,vidal2003entanglement,gokmen2021statistical}.
The mutual information is particularly well-suited to extract the correlations between two open subsystems, as it cancels out contributions to the bipartite entropy that arise from entanglement with the rest of the system \cite{Vedral2001,Modi2010,Artiaco_PRL2025}. 
This makes it preferable to the bare von Neumann entropy \cite{neumann1955mathematical,Calabrese_EEreview}, which has a clear meaning only
for pure states.

In this work, we specify these ideas to our multi-component system by focusing on the local \emph{inter-flavor} mutual information, a rather general measure of correlations between two out of the $N$ flavors of the model. For the SU(2) case, this quantity has already been introduced for the two spin species as  \emph{intra-orbital} mutual information and it has been found to faithfully account for the correlation properties of Hubbard systems 
\cite{Bellomia_intracorr,bellomia2024quantum}. Additionally, for systems where each fermionic flavor is individually conserved, the inter-flavor mutual information depends only  on experimentally accessible quantities, such as average double occupancies and flavor-resolved densities. 

\subsection{\label{sec: entropy} Two-flavor and single-flavor entropy}

The main target of our analysis are the local correlations between pairs of flavors. Therefore, we start from the local two-flavor reduced density matrix $\rho^{(i)}_{\alpha\beta}\equiv\rho_{\alpha\beta}$, obtained  tracing out all lattice sites except site $i$ and all flavors different from $\alpha$ and $\beta$. In the absence of inhomogeneities and spatial symmetry breaking, we can drop the lattice site label $i$. From this we compute the \emph{two-flavor entropy}
\begin{equation}\label{eq: 2-RDM}
    s_{\alpha\beta} = -\text{Tr}\,\rho_{\alpha\beta}\,\text{log}\rho_{\alpha\beta}.
\end{equation}
The local configurations of two flavors are simply $\{ |0\rangle, |\alpha\rangle, |\beta\rangle, |\alpha,\beta\rangle \}$. Since the Hamiltonian individually conserves the number of particles for each flavor (in the basis of natural orbitals), the two-flavor density matrix is diagonal
\cite{zanardi2002quantum,Schilling_QuantumScience,Bellomia_intracorr} and 
it takes the form 
\begin{equation}
    \label{eq: rho_ab}
    \rho_{\alpha\beta} = \text{diag}\left( p_0, p_\alpha, p_\beta, p_{\alpha,\beta} \right),
\end{equation}
where $p_0, p_\alpha, p_\beta, p_{\alpha,\beta}$ represent the probabilities of the site being empty, singly occupied by flavor $\alpha$, singly occupied by flavor $\beta$, and occupied by both flavors, respectively \cite{walsh2019local,Held_RDMs,bellomia2024quantum}, regardless of the other flavors. One has $p_{\alpha,\beta}=\langle n_\alpha n_\beta \rangle$, $p_\alpha=\langle n_\alpha \rangle - \langle n_\alpha n_\beta \rangle$, $p_\beta=\langle n_\beta \rangle - \langle n_\alpha n_\beta \rangle$ and $p_0=1 - \langle n_\alpha \rangle - \langle n_\beta \rangle + \langle n_\alpha n_\beta \rangle$. Hence, the two-flavor density matrix, as well as all the entropic quantities that depend on it, can be easily computed in terms of observables accessible in cold atom experiments \cite{flavor_selective_experiment}, providing a much simpler measure of correlations with respect to those based on the knowledge of the full $N$-flavor density matrix.


Finally, $\rho_{\alpha\beta}$ can be partitioned into its single-flavor components, which we denote $\rho_{\alpha}$ and $\rho_{\beta}$ respectively. These are as well diagonal and take the form $\rho_{\sigma}=\text{diag}\left(\langle n_\sigma \rangle, 1-\langle n_\sigma \rangle\right)$ for $\sigma \in \{ \alpha, \beta \}$. The single-flavor reduced density matrices directly yield the single-flavor entropies 
$s_\alpha = -\text{Tr}\,\rho_\alpha\,\text{log}\rho_\alpha$ and 
$s_\beta= -\text{Tr}\,\rho_\beta\,\text{log}\rho_\beta$.

\subsection{\label{sec: mutualinfo} Inter-flavor mutual information}
We can now define the von Neumann mutual information between $\alpha$ and $\beta$
\begin{equation}
    \label{eq: I}
    I_{\alpha\beta}=s_{\alpha}+s_{\beta}-s_{\alpha\beta},
\end{equation}
which we call \emph{inter-flavor mutual information}. This quantity captures both quantum and classical correlations between the two subsystems, as it follows from a clean geometrical picture \cite{Vedral_RMP,Plenio_EntanglementMeasures,Schilling_QuantumScience,bellomia2024quantum}. In this framework, correlations are quantified by the \emph{informational distance} between $\rho_{\alpha\beta}$ and the closest uncorrelated state $\sigma = \sigma_\alpha\otimes\sigma_\beta$. A widely adopted choice for this distance is the \emph{quantum relative entropy} $S(\rho||\sigma)$, defined as
\begin{equation}
    \label{eq: QRE}
    S(\rho||\sigma) = \mathrm{Tr}\,\rho\log\rho - \mathrm{Tr}\,\rho\log\sigma,
\end{equation}
which is zero if and only if $\rho=\sigma$, as expected for a well-behaved metric \cite{Vedral2001}. The quantum relative entropy thus
quantifies how distinguishable the states $\rho$ and $\sigma$ are. It was proven in \cite{Vedral_RMP} that the closest uncorrelated state to $\rho_{\alpha\beta}$ is precisely $\rho_\alpha\otimes\rho_\beta$, and that the corresponding minimal distance measured by the quantum relative entropy is $I_{\alpha\beta}$. More clearly, one can rewrite the mutual information in terms of the quantum relative entropy as
\begin{equation}
    I_{\alpha\beta}= \min_{\{\sigma_\alpha,\sigma_\beta\}}
    S(\rho_{\alpha\beta}||\sigma_\alpha\otimes\sigma_\beta) =
    S(\rho_{\alpha\beta}||\rho_\alpha\otimes\rho_\beta).
\end{equation}
The mutual information is a good marker for correlations, since it shows how far the composite state $\rho_{\alpha\beta}$ is from the uncorrelated product state $\rho_\alpha\otimes\rho_\beta$.
In our case, the indices $\alpha$ and $\beta$ label the local \emph{natural} orbitals \cite{Natural_spin-orbitals_def} of the Hamiltonian of \cref{sec: model}. This uniquely defined bipartition minimizes correlations across all possible choices,
so $I_{\alpha\beta}$ captures the genuine degree of correlation in the composite
${\alpha\beta}$ system \cite{Faraday_Natural_Orbitals}. 
Moreover, it has been recently proven that the mutual information between natural orbitals can be explicitly connected to the concept of \emph{fermionic nongaussianity} \cite{Gottlieb2005,Gottlieb2007,Paris_nonGaussianity,GeometricEntropy_MIT_Vollhardt,Held_nonfreeness,LyuBu24_nonGaussianity,Sierant2025_FAF,Sierant2026_FAF,Matchgate_Review,Mele_NonfreenessBounds,Mele_NonfreenessLearning,Lami2025_nongauss_doping}, measured as the minimal relative entropy from the set of free fermion states,
clarifying that $I_{\alpha\beta}$ measures correlations between physical particles, as they visit the $\alpha\beta$ system \cite{Bellomia_intracorr}.
Furthermore, $I_{\alpha\beta}$ is an upper bound for correlation functions \cite{wolf2008area}
\begin{equation}
    \label{eq: I_bound}
    I_{\alpha\beta} \geq \frac{\left(\langle\mathcal{O}_\alpha \otimes \mathcal{O}_\beta\rangle_{\rho_{\alpha\beta}} - \langle\mathcal{O}_\alpha\rangle_{\rho_\alpha}\langle\mathcal{O}_\beta\rangle_{\rho_\beta}\right)^2}{2||\mathcal{O}_\alpha||^2||\mathcal{O}_\beta||^2},
\end{equation}
where $\mathcal{O}_\alpha$ and $\mathcal{O}_\beta$ are generic operators acting on the Hilbert spaces of $\alpha$ and $\beta$ respectively, and their norms are defined as the respective maximal singular values.

In order to discriminate quantum and classical correlations we consider the ensemble of \emph{separable} (i.e.\,unentangled) states and its subset of \emph{pseudo-classical} states (i.e.\,states whose correlations are effectively classical \cite{Discord_def,Vedral2001,Modi2010,Schilling_QuantumScience}). 
Following the geometrical approach, entanglement $E_{\alpha\beta}$ can be quantified by the distance between $\rho_{\alpha\beta}$ and the closest separable state, all quantum correlations $Q_{\alpha\beta}\geq E_{\alpha\beta}$ are captured by the distance from the closest pseudo-classical state, while a measure of classical correlations $C_{\alpha\beta}$ can be defined as the distance between the closest pseudo-classical state and the closest uncorrelated state. These geometric quantifiers of correlations satisfy the chain of inequalities \cite{Schilling_QuantumScience}.
\begin{equation}
    I_{\alpha\beta} \geq C_{\alpha\beta} + Q_{\alpha\beta} \geq C_{\alpha\beta} + E_{\alpha\beta}
\end{equation}
For pure states quantum correlations are fully determined by entanglement, as measured by the bipartite von Neumann entropy 
$Q_{\alpha\beta}=E_{\alpha\beta}=s_\alpha=s_\beta$. 
On the other hand, for pseudo-classical states \cite{Discord_def,Vedral2001,Modi2010,Schilling_QuantumScience}, $Q_{\alpha\beta}=0$ and $I_{\alpha\beta}=C_{\alpha\beta}$, so that the mutual information is a measure of classical correlations in the system. 
As the Hamiltonian in \cref{eq: H} conserves the number of flavors on the whole lattice, the symmetry-enforced diagonal form of the two-flavor reduced density matrix (\cref{eq: rho_ab}) ensures that the quantum state of any given pair of flavors residing on a single lattice site is pseudo-classical \cite{Bellomia_intracorr,bellomia2024quantum} and so all the inter-flavor correlations are classical.
Hence, in this setup, the degree of local nongaussianity constitutes a resource of state complexity that is fully independent from local quantum correlations and entanglement.

\section{\label{sec: results} Results}
In this section, we present our DMFT results.
In \cref{sec: su2vssu4}, we focus on the SU(4) symmetric Hubbard model, highlighting its differences and analogies with the SU(2) model. In order to further explore the role of the number of components, in \cref{sec: sun} we study the SU($N$) symmetric model in the atomic limit, where we can vary $N$ easily. Finally, in \cref{sec: omega}, we study the  breaking the SU(4) symmetry of \cref{eq: H} by allowing $\Omega \neq 0$.
\subsection{\label{sec: su2vssu4} SU(4) vs SU(2) Mott transitions}

\begin{figure}
  \includegraphics[width=\linewidth]{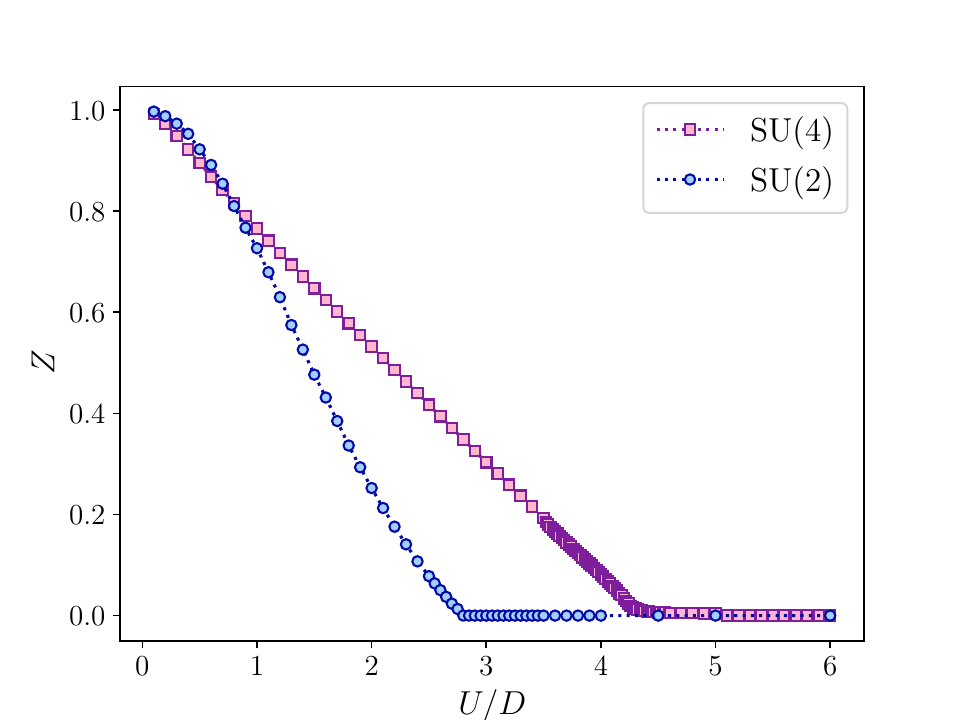}
  \caption{Quasiparticle weight $Z_\alpha$ for the SU(2) and SU(4) Hubbard models as a function of $U/D$. The vanishing of $Z_\alpha$ marks the Mott transition.
  }
  \label{fig: z_sun}
\end{figure}

\begin{figure}
  \includegraphics[width=\linewidth]{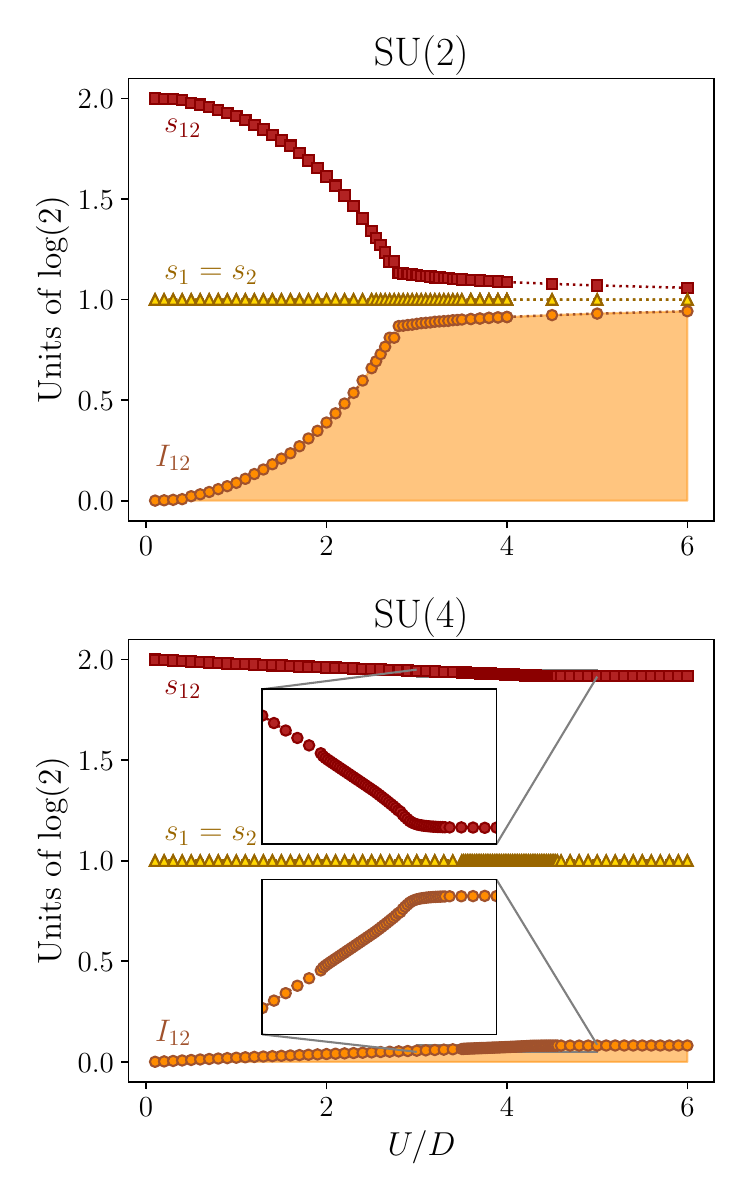}
  \caption{Two-flavor entropy $s_{12}$, single-flavor entropy $s_1=s_2$ and inter-flavor mutual information $I_{12}$, plotted against the interaction strength, for the SU($2$) Hubbard model (top panel) and the SU($4$) model (bottom panel). The area under the mutual information is shaded to indicate that all local correlators are contained beneath it.  All quantities are in units of $\log(2)$ (bits, in information theory language).
  }
  \label{fig: su2su4}
\end{figure}
We begin by comparing the fully symmetric SU(4) model with the more conventional SU(2) system.
The metal-insulator transition is signaled by the flavor-resolved quasiparticle weight, defined as
\begin{equation}
    \label{eq: z}
    Z_\alpha=\left( 1 - \left.\frac{\partial\text{Im}\Sigma_{\alpha\alpha}(i\omega)}{\partial i\omega}\right|_{\omega=0}\right)^{-1},
\end{equation}
with $\Sigma_{\alpha\alpha}(i\omega)$ the momentum-independent diagonal self-energy matrix. Here, because of the SU($N$) symmetry, $Z_\alpha$ does not depend on $\alpha$. $Z_\alpha$ measures the weight of the quasiparticle excitations, and it equals 1 in a noninteracting metal, while it vanishes in a Mott-Hubbard insulating state.  Our results, shown in Fig.\,\ref{fig: z_sun}, confirm previous literature and find a continuous evolution of  $Z_\alpha$ before reaching a critical interaction  $U/D \simeq 2.8$ for the SU(2) case and a significantly larger value of $U/D \simeq 4.2$ for SU(4). Besides this quantitative difference, the two transitions seem similar based on these results.

The mutual information between two flavors, presented in \cref{fig: su2su4}, 
reveals instead a  completely different picture in SU(2) and SU(4) systems. Indeed, while the SU(2) case has $I_{12}$ continuously approaching a value close to $\log{2}$ at the Mott transition,
the same quantity appears always small for the SU(4) system. Yet, if we zoom in, we see that also in this case $I_{12}$ is a monotonically increasing function of $U/t$, but it reaches a maximum value of $0.08 \log{2}$, signaling significantly lower correlations in the Mott insulator compared to the SU(2) case. The much smaller value descends from the fact that $s_{12}$ remains close to $2 \log{2}$ as we increase $U/t$, in contrast with the SU(2) case, where it drops from $2 \log{2}$ to $\log{2}$. Our results show that the behavior of $Z$ hides a profound difference between the SU(2) and the SU(4) systems. In the next section, we discuss the effect of further increasing $N$ in the atomic limit, where the model can be solved for arbitrary $N$.

\subsection{\label{sec: sun} Atomic  SU($N$) model at any value of $N$}

\begin{figure}
  \includegraphics[width=\linewidth]{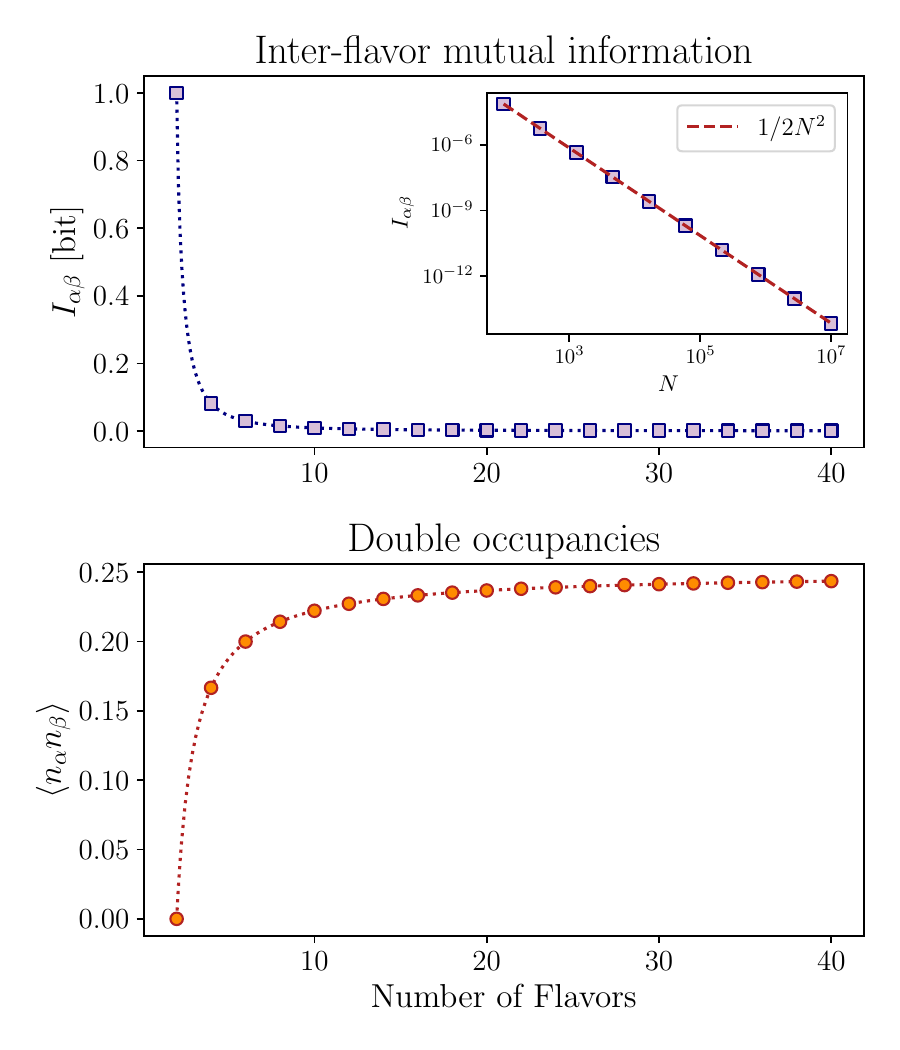}
  \caption{Inter-flavor mutual information (top) and double occupancy per site (bottom) for a half-filled SU($N$) Hubbard model in the atomic limit $t=0$, plotted against the total number of flavors $N$. 
  In the inset we show again the mutual information vs. $N$ in a log-log plot highlighting the $1/2N^2$ scaling (red dashed line). 
  }
  \label{fig: sun}
\end{figure}

\begin{figure*}[t]
  \centering
  \begin{minipage}[t]{0.49\textwidth}
    \centering
    \includegraphics[width=\linewidth]{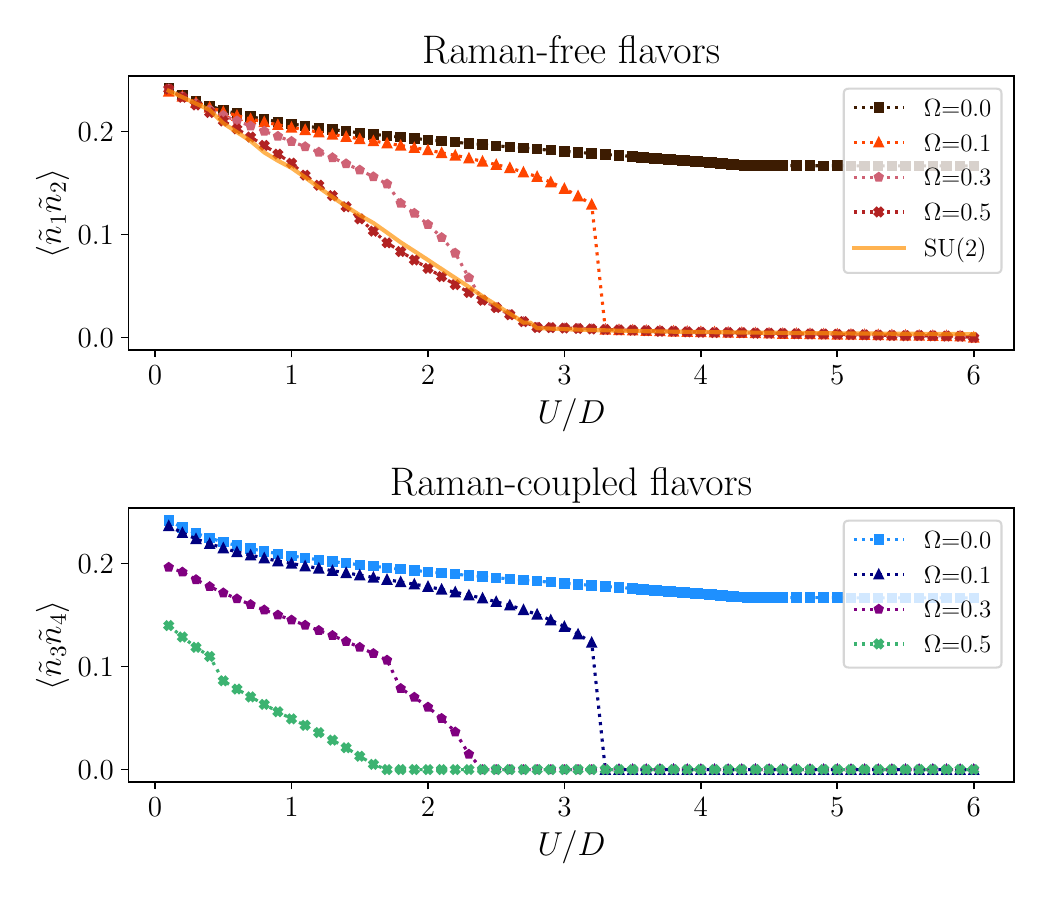}
    \caption{Double occupancies per site for the Raman-free flavors (top panel) and the Raman-coupled flavors (bottom panel) for different values of the Raman coupling $\Omega$. For Raman-coupled flavors, double occupancies vanish for $U > U^{\mathrm{R}}_{\mathrm{c}}(\Omega)$ due to the occurrence of complete polarization, while the Raman-free ones display a crossover between the values in the SU($4$) and SU($2$) Hubbard model.}
    \label{fig:docc}
  \end{minipage}
  \hfill
  \begin{minipage}[t]{0.49\textwidth}
    \centering
    \includegraphics[width=\linewidth]{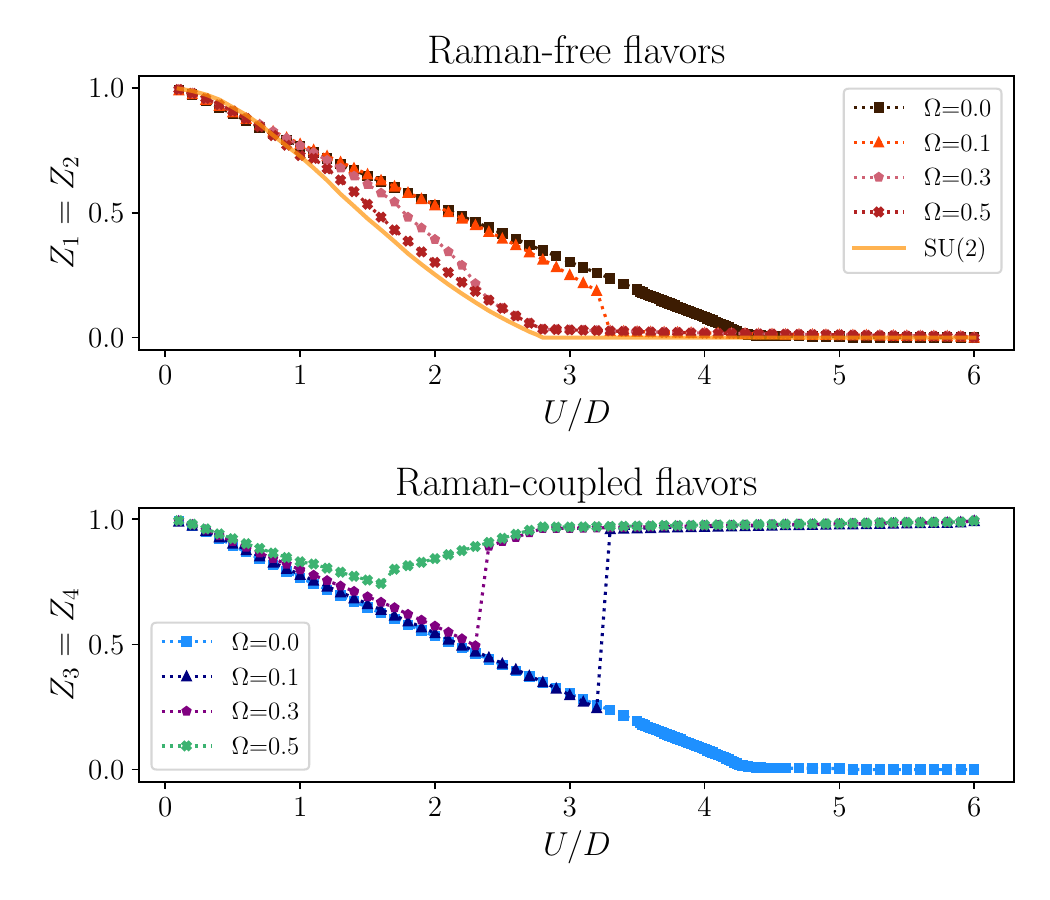}
    \caption{Quasiparticle weight for the Raman-free flavors (top panel) and the Raman-coupled flavors (bottom panel) for different values of the Raman coupling $\Omega$. For Raman-coupled flavors, the quasiparticle weight tends to $1$, signaling the hybrid nature of the insulating state, while the Raman-free $Z$ vanishes for $U > U^{\mathrm{M}}_{\mathrm{c}}(\Omega)$, marking the onset of the Mott metal-to-insulator transition.}
    \label{fig:z}
  \end{minipage}
\end{figure*}

In order to extend results to large values of $N$, for which a DMFT calculation would rapidly become unfeasible, we consider the atomic limit  $t=0$ where the model simplifies to a collection of disconnected Hubbard atoms. 
The density of each flavor is $\langle n_\alpha \rangle = n/N$, where $n$ is the number of fermions per site, while the double occupancies are given by the ratio between all possible couples among $n$ particles and all possible couples among $N$ flavors, i.e.\,$\langle n_\alpha n_\beta \rangle = \frac{n(n-1)}{N(N-1)}$ for every $\alpha\neq \beta$.

In \cref{fig: sun} we display how the inter-flavor mutual information and the double occupancies evolve with the number of flavors at half filling ($n=N/2$). 
The mutual information decreases rapidly with the number of flavors, leading to very weakly correlated Mott insulators as $N$ becomes large. Hence, in this limit the Mott state, despite being driven by large interactions,  lacks inter-flavor correlations. Consequently, since the mutual information is an upper bound to all correlation functions, \cref{eq: I_bound} implies that, in the large-$N$ limit, the correlation function of any pair of local operators must vanish at least as fast as $I_{\alpha\beta}$.
, approaching an uncorrelated mean-field description, with $\langle \mathcal{O}_\alpha \mathcal{O}_\beta \rangle \to \langle \mathcal{O}_\alpha \rangle \langle \mathcal{O}_\beta \rangle,\, \forall \mathcal{O}_\alpha, \mathcal{O}_\beta$. 
We show the case of double occupancies, that approach their factorized value $\langle n_\alpha n_\beta\rangle \to \langle n_\alpha\rangle\langle n_\beta\rangle=1/4$ for $N\rightarrow\infty$, as shown in the bottom panel of \cref{fig: sun}. Our analytical treatment gives thus a strong indication that, in the atomic limit, the local correlations approach a mean field behavior,  where the flavors are decoupled. Furthermore, at half filling it can be analytically proven that the leading term for the mutual information at large $N$ scales as $1/2N^2$, as confirmed by the inset of \cref{fig: sun}. It is interesting to observe that the drop from the large value of $N=2$ to small values is very rapid, and already $N=4$ has a remarkably smaller value of $I_{12}$. In this light the standard two-species Mott transition appears to be intrinsically different from any other value of $N$. We finally mention that the trends are similar also for any other integer filling $n \neq N/2$.

\subsection{\label{sec: omega} Broken SU(4) symmetry}

\begin{figure*}
  \includegraphics[width=\textwidth]{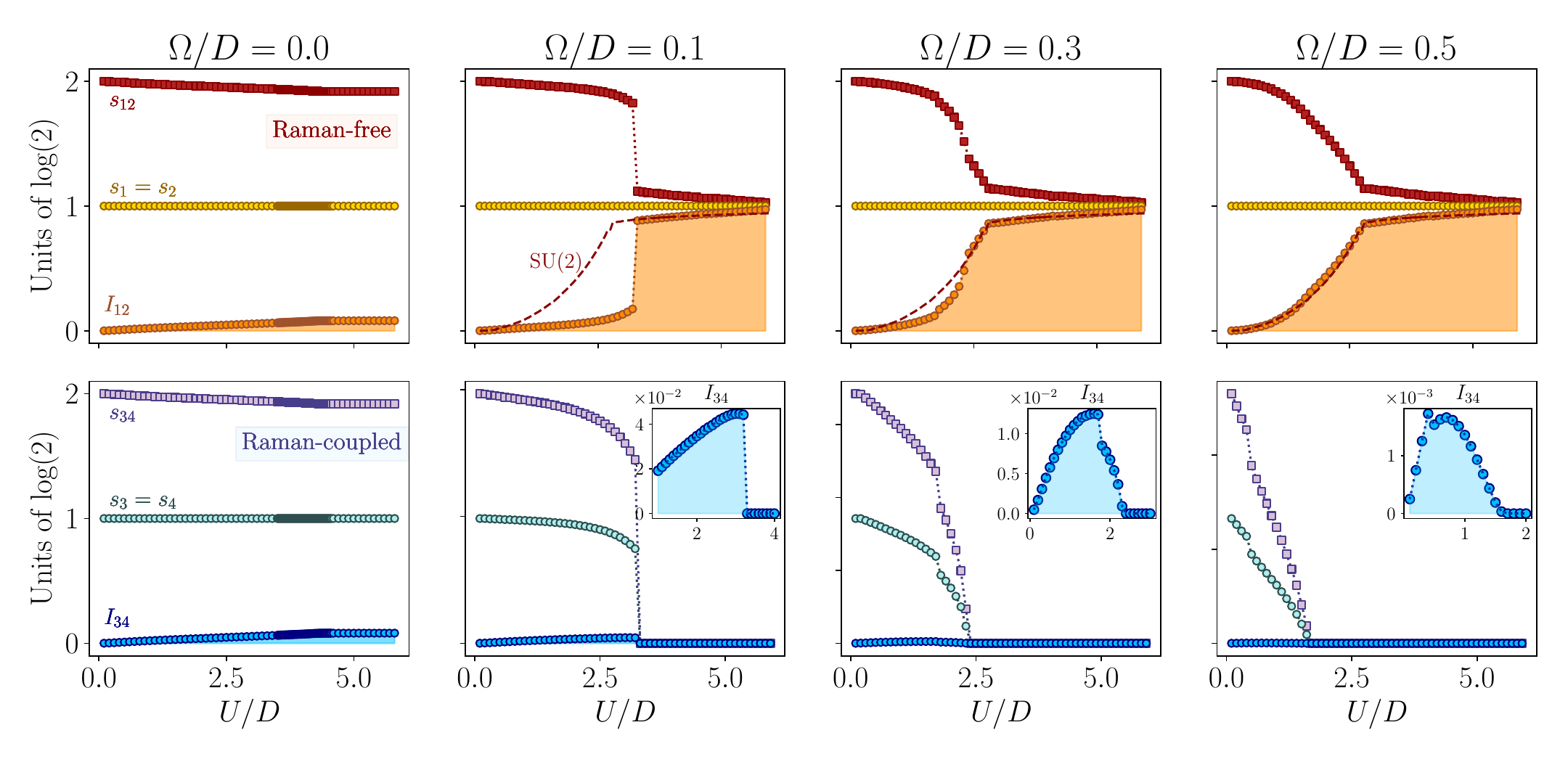}
  \caption{Two-flavors entropy $s_{\alpha\beta}$, single-flavor entropy $s_\alpha, s_\beta$ and inter-flavor mutual information $I_{\alpha\beta}$, plotted against the interaction strength $U$, for different values of the Raman coupling $\Omega$, indicated on top of each column, for the Raman-free flavors (top panel) and the Raman-coupled flavors (bottom panel). The insets in the bottom panel are close-ups of the mutual information between Raman-coupled flavors, highlighting that it vanishes for $U>U^{\mathrm{R}}_{\mathrm{c}}(\Omega)$. Raman-free flavors display a clear crossover between the SU($2$) and the SU($4$) ($\Omega=0$) model as $\Omega$ increases. All quantities are in units of $\log(2)$.}
  \label{fig: entropies_omega}
\end{figure*}

 We now include a finite $\Omega\neq 0$. As a result the original SU($4$) symmetry, supplemented by a global charge conservation is reduced to $SU(2) \otimes U(1) \otimes U(1)$, where SU($2$) is the usual symmetry for the Raman-free flavors, and the two U($1$) are generated by the conservation of $\tilde{N}_{3} - \tilde{N}_{4}$ and $N_1+N_2- \tilde{N}_{3} - \tilde{N}_{4}$ respectively.

The effect of a finite $\Omega$ is to shift the energies of the Raman-coupled fermions in opposite directions, thus creating a population imbalance between the two species. The Raman-free levels are instead unaffected and they remain in the middle of the two coupled levels.

We anticipate that the effect of an increased $U$ is to effectively enhance such energy separation, eventually opening a gap between the bands of the Raman-coupled flavors and leading to localization, similarly to what has been predicted and observed in the SU(3) case \cite{del2018selective,flavor_selective_experiment}.

As a result, for each value of $\Omega$ there is a critical value of the interaction, which we denote $U_{\mathrm{p}}\left(\Omega\right)$, at which the Raman-coupled flavors become completely polarized, namely $\tilde{n}_4-\tilde{n}_3 = 1$ for all values of $U>U_{\mathrm{p}}\left(\Omega\right)$. 
Thus, the Raman coupled bands do not display Mott localization, which is confined to the  SU(2)-symmetric Raman-free flavors, which is always half-filled. This leads to a crossover between the SU(4) and the SU(2) Mott transition.
The sharp difference between the Raman-free and the Raman-coupled flavors signals \emph{flavor-selective} behavior, in which different flavors undergo different fates.

The evolution of the average double occupancies per site across the Mott transition, displayed in \cref{fig:docc}, reflects the above-mentioned polarization of the Raman-coupled flavors, as well as the SU($4$)-SU($2$) crossover for the Raman-free ones. For the latter case, shown in the top panel of \cref{fig:docc}, when $\Omega=0$ the double occupancies decrease and approach $1/6$ in the limit $U \gg D$, as expected for a SU($4$) Mott insulator. In contrast, for any $\Omega>0$, double occupancies are suppressed in the Mott insulating phase, aligning with the SU($2$) behavior as the coupling with the Raman field increases. This clearly highlights the presence of a crossover between the two models driven by the tuning of $\Omega$. 
On the other hand, the double occupancies of the Raman-coupled flavors, shown in the bottom panel of \cref{fig:docc}, display clear signs of growing polarization between the two flavors as $U$ increases. When $U$ increases beyond $U_p\left(\Omega\right)$, the double occupancies drop to zero, signaling that the Raman-coupled flavors have become fully polarized, with only one of their two bands occupied.

As shown in the top panel of \cref{fig:z}, the quasiparticle weight of the Raman-free flavors vanishes at $U_c\left(\Omega\right)$, marking the transition into the Mott insulating phase. 
For every $\Omega$, $Z$ lies between its values observed in the SU($4$) and in the SU($2$) model, approaching the SU($2$) curve as $\Omega/D$ grows closer to $1$. Indeed, for these flavors the critical interaction strength for the metal-to-insulator transition decreases from $U_{\mathrm{c}} \simeq 4.2$, when $\Omega=0$ (characteristic of the SU($4$) model) to $U_{\mathrm{c}}, \simeq 2.8$, when $\Omega\simeq D$, which is precisely the critical value observed in the SU($2$) model.
On the contrary, the quasiparticle weight of the Raman-coupled flavors tends to 1 for large values of $U/D$, for all $\Omega>0$, as shown in the bottom panel of \cref{fig:z}. This reflects the hybrid nature of the large-$U$ state, which is not a fully incoherent Mott insulator but has the Raman-coupled flavors behaving as in a band insulator, where quasiparticles remain well-defined\footnote{For the Raman-coupled flavors, $Z$ was not computed using \cref{eq: z}, but rather as the residue of the pole of the Green's function at $\omega^*$, determined by the condition $\omega^*-\varepsilon-\text{Re}\Sigma(\omega^*)=0$. Due to the locality of the self-energy, $Z$ is independent on $\varepsilon$
, therefore we compute it for $\varepsilon=0$.
This approach is made necessary because, in a band insulator, quasiparticles are not located at the Fermi level but in the conduction and valence bands.}.

Importantly, we can study how inter-flavor correlations evolve with the symmetry breaking term, as shown in \cref{fig: entropies_omega}.
At $\Omega=0$, when $U>U_{\mathrm{c}}\left(0\right)$ the system is a weakly correlated SU($4$) Mott insulator, as already discussed in \cref{sec: su2vssu4} (see \cref{fig: su2su4}). 
As $\Omega$ increases, correlations between Raman-free flavors in the insulating phase become significantly stronger, eventually matching the values observed in the SU($2$) model, as shown in the top row of \cref{fig: entropies_omega}. 
It is then clear how reducing the symmetry between flavors leads to an enhancement of local inter-particle correlations, recovering strong SU(2)-like correlations in the Mott-insulating state.

On the other hand, the mutual information between Raman-coupled flavors vanishes when the two flavors become polarized, i.e.\,for $U>U_{\mathrm{c}}\left(\Omega \right)$, as a consequence of the fact that one of the two bands is empty and no correlation can take place. The vanishing of the mutual information emphasizes that the transition is driven by the polarization-induced gap, rather than a Mott-Hubbard mechanism, similarly to what has been proposed for maximally magnetized antiferromagnets \cite{Bellomia_intracorr} and for synthetic systems with a gauge field \cite{ferraretto_scipost}.

The evolution of entropy-based quantities not only neatly captures the crossover between the SU($4$) and the SU($2$) Mott-Hubbard transitions, but also illustrates how inter-particle correlations in the Mott phase can be effectively tuned  by controlling the symmetry of the model. Remarkably, one can recover a strongly-correlated insulator from a multi-component system by breaking the underlying flavor symmetry.

\subsection{Phase diagram and the tricritical point}


\begin{figure*}[t] 
\centering
    \includegraphics[trim={0.5cm 0 1.5cm 0},clip,width=.48\linewidth]{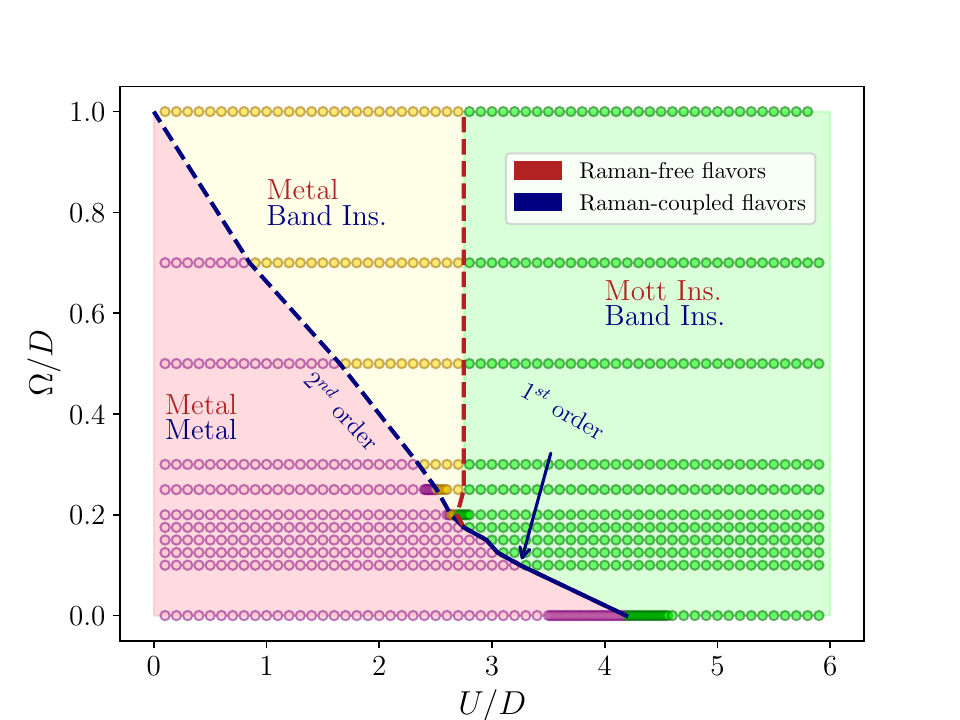}
    \includegraphics[trim={0 0 0.5cm 0},clip,width=.5\linewidth]{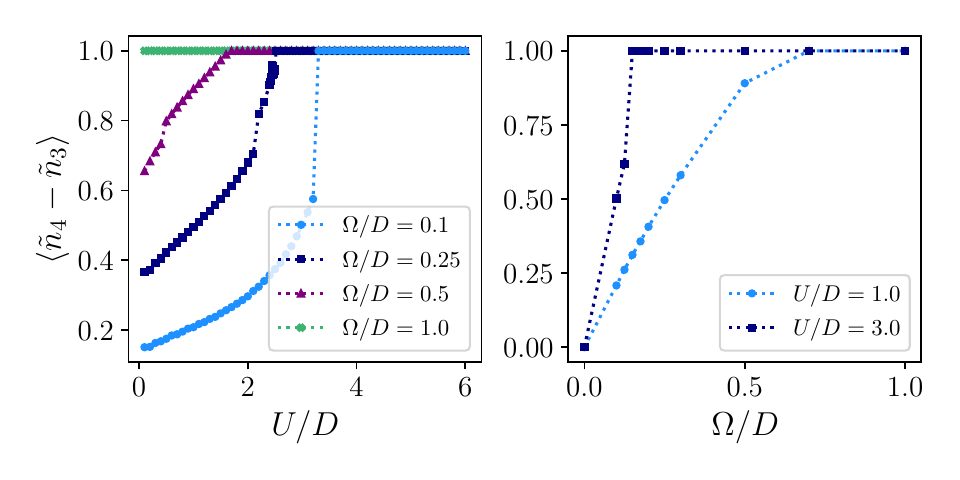}
    \caption{Phase diagram in the $\Omega/D$-$U/D$ plane, showing the different competing phases. The dashed red line, $U_{\mathrm{c}}\left(\Omega\right)$, marks the Mott transition of the Raman-free flavors, while the dashed blue line, $U_{\mathrm{p}}\left(\Omega\right)$, represents the polarization transition of the Raman-coupled flavors. These lines intersect at the tricritical point $\left(U^\mathrm{T}/D,\Omega^\mathrm{T}/D\right)$, where the transition of the Raman-coupled flavors changes from second to first order. The right panel displays the polarization of the Raman-coupled flavors as a function of $U/D$ (left) and $\Omega/D$ (right). For $\Omega<\Omega^\mathrm{T}$, the two dashed lines merge into a continuous one, indicating that the Raman-free and Raman-coupled flavors undergo their phase transition simultaneously.
    }
    \label{fig: phase_diag}
\end{figure*}

We summarize our findings in the phase diagram of \cref{fig: phase_diag} (upper panel). In general, the system exhibits three distinct phases: A metallic phase in which all flavors have itinerant character; a flavor selective state in which the Raman-coupled flavors are in a a band insulating state due to the polarization-induced gap, while the Raman-free ones remain metallic, and a Mott insulating state where all the flavors are localized by the interaction.

The transition between these phases is determined by the two critical values of $U$. 
The red line in \cref{fig: phase_diag} represents $U_{\mathrm{c}}\left(\Omega\right)$ and marks the Mott transition of the Raman-free flavors, signaled by the vanishing of the quasiparticle weights $Z_1$, $Z_2$. 
The blue line represents $U_{\mathrm{p}}\left(\Omega\right)$ and determines the critical points where the Raman-coupled flavors become fully polarized, meaning that $\tilde{n}_4-\tilde{n}_3=1$, and form a band insulator. 
These two transition lines appear to intersect at a tricritical point, $\left(U^\mathrm{T},\Omega^\mathrm{T}\right)$, where all three phases coexist.
At this point, the nature of the transition for the Raman-coupled flavors changes. For $\Omega<\Omega^\mathrm{T}$, the transition is found to be clearly first order, characterized by a discontinuous jump in the flavor polarization $\tilde{n}_4-\tilde{n}_3$, while for $\Omega>\Omega^\mathrm{T}$, the transition appears of second order within our numerical accuracy, with the polarization smoothly approaching $1$, i.e.\,the value defining a complete polarization. 
This difference is illustrated in the bottom panel of \cref{fig: phase_diag}, where both horizontal and vertical cuts through the phase diagram show the polarization behavior. Additionally, for $\Omega<\Omega^\mathrm{T}$, the two transition lines merge, indicating that both Raman-free and Raman-coupled flavors undergo the phase transition simultaneously.

These observations suggest that $(U^\mathrm{T},\Omega^\mathrm{T})$ behaves like a classical tricritical point, where three phases coexist and the first- and second-order phase transitions intersect \cite{landau_tricritical,landau_statphys_book}. Our DMFT analysis places this point at $\Omega^\mathrm{T}\simeq0.2D$ and $U^\mathrm{T}\simeq2.8D=U^{\text{SU(2)}}_c$, which corresponds to the critical interaction strength for the SU(2) Mott transition. This result is yet another consequence of the SU(4)-SU(2) crossover induced by $\Omega$. Indeed, for $\Omega>\Omega^\mathrm{T}$, the transition becomes purely SU(2)-like, thus the quasiparticle weight of the Raman-free flavors vanishes always at $U _c(\Omega)=U^{\text{SU(2)}}_c$ and the critical interaction does not depend on $\Omega$.

\section{\label{sec: conclusions} Conclusions}

\subsection*{\!\!\!\!\!Summary}
In this work, we studied the degree of correlation between flavors in the SU($N$) Hubbard model in the geometrical framework associated to quantum relative entropies, an ideal suit for open quantum subsystems \cite{Vedral_RMP,Plenio_EntanglementMeasures,EntanglementDetection_review,Schilling_QuantumScience,bellomia2024quasilocal,bellomia2024quantum}, such as fermionic orbitals embedded in a lattice. By recognizing that the symmetries of the model constrain all local (i.e.\,residing on a single-site) reduced density matrices to be pseudo-classical \cite{Discord_def}, we ensured that neither entanglement nor quantum discord is shared by the local flavors, so that the inter-flavor mutual information gives a faithful measure of local classical correlations between particles \cite{Bellomia_intracorr}, relating to the concept of fermionic nongaussianity \cite{Gottlieb2005,Gottlieb2007,Paris_nonGaussianity,Held_nonfreeness,LyuBu24_nonGaussianity,Sierant2025_FAF,Sierant2026_FAF,Matchgate_Review,Mele_NonfreenessBounds,Mele_NonfreenessLearning,Lami2025_nongauss_doping,GeometricEntropy_MIT_Vollhardt}.
We characterized the behavior of this quantity across the Mott-Hubbard transition for $N=2$ and $N=4$ and examined the effects of a tunable symmetry breaking term in the latter case, effectively uncovering a rich crossover between SU(4) and SU(2)-symmetric case.
We benchmarked the behavior of the inter-flavor mutual information, a quantity accessible to cold atom experiments, by comparing it to the evolution of the quasiparticle weight, an experimentally challenging quantity, naturally linked to Fermi liquid theory and its breakdown at genuine Mott-Hubbard transitions.

In the absence of symmetry breaking, we showed how classical local correlations monotonically increase with the inter-flavor interaction strength $U$, eventually saturating in the Mott-Hubbard insulating phase. 
However, the number of available flavors in the lattice significantly affects the degree of correlation of the insulator, which becomes less correlated as $N$ increases. We formalized this statement in the atomic limit, where the hopping is set to zero and the model, reduced to a collection of Mott-Hubbard states, is analytically solvable.
In this regime, we computed the mutual information between flavor pairs, and found that the correlations vanish in the large-$N$ limit. It follows that in the presence of a large number of flavors, the ``correlation-driven" Mott insulator becomes effectively uncorrelated. 
Additionally, we showed that the vanishing of the inter-flavor mutual information can be characterized analytically at half-filling, as $\mathcal{O}(1/2N^2)$ for $N\rightarrow\infty$. Since the mutual information is an upper bound for all correlation functions between pairs of local operators, its vanishing implies that all local expectation values factorize in the large-$N$ limit. This is consistent with the evolution of local double occupancies, which approach their factorized values when $N\gg 1$, i.e.\,$\langle n_\alpha n_\beta \rangle \rightarrow \langle n_\alpha \rangle \langle n_\beta \rangle$, confirming the emergence of mean-field behavior.
This apparently counter-intuitive decoupling is actually a well-explored feature of Yang-Mills theories, namely gauge theories based on the group SU($N$), where it was shown that in the large-$N$ limit quantum fluctuations are damped and mean-field-like correlations emerge \cite{thooft_largeN,tong_gauge}. Yet, we highlight the striking quantitative distance between the SU(2) and the SU(4) scenarios, with the latter effectively closer to the large-$N$ limit than to the highly singular SU(2) model.

To further investigate the behavior of correlations in the small-$N$ regime, we showed how introducing a suitable Raman term in the $N=4$ case induces a crossover between the SU($4$) and the SU($2$)-symmetric Hubbard model for the Raman-free flavors, while the Raman-coupled ones become completely polarized. 
This provides a remarkable example of flavor-selective behavior: the Raman-free flavors undergo an SU($2$) Mott transition, whereas the Raman-coupled ones form an uncorrelated band insulator. The inter-flavor mutual information neatly captures this crossover, and provides clear-cut insight into the contrasting nature of Raman-coupled and Raman-free flavors. Thus, we reinforce previous evidence that this quantity, as well as closely related geometrical measures of entanglement and quantum correlations, represent a crucial tool not only for the generic detection and characterization of critical behavior and phase transitions, but also for the classification of fermionic states in terms of strong, interaction-induced, correlations \cite{Bellomia_intracorr,bellomia2024quasilocal,bellomia2024quantum}.

\subsection*{\!\!\!\!\!Outlook to experiments: \\ cold atoms and solid state counterparts}
We conclude by briefy discussing how this model, and the tuning from SU(4) to SU(2) are readily realizable in experiments with ultracold atoms in optical lattices, which have repeatedly proven their role as essential platforms to probe the physical properties of SU($N$) fermions \cite{SUN_experiment_2,SUN_experiment_3,SUN_experiment_4,SUN_experiment_5,SUN_experiment_6,ultracold_SU(N),scazza_SU(N),alkaline_SU(N),flavor_selective_experiment,Scazza_SU(N)_crossover,Bloch_SU(N)}.
The set-up that we propose can be realized via a simple extension of that of Ref.\,\cite{flavor_selective_experiment}: starting, for instance, from fermionic $^{173}$Yb atoms, one can filter an initial state with equal number of fermions in four nuclear spin components, realizing a SU(4) system which can be loaded in a a three-dimensional optical lattice. The SU(4) symmetry can be broken as in our model by two-photon Raman processes  with Rabi frequency $\Omega/h$ (where $h$ is the Planck constant) connecting two flavors and leading to dressed linear combinations.
In order to provide experimental evidence, one should measure flavor-selective occupations and double occupations (or correlators) of the form $\langle n_{\alpha}n_{\beta} \rangle$. 
The double occupancy has been measured in \cite{flavor_selective_experiment} via one-color photoassociation  processes that act only when an atom is in a doubly-occupied lattice site. A small magnetic field allows to make the measurement sensitive to the flavors, providing us with flavor-selective quantities and, consequently, with a measurement of the inter-flavor mutual information. 

SU($N$) symmetric models of interacting fermions have also been proposed as simplified descriptions of complex layered materials, like moiré transition-metal dichalcogenides \cite{su4_moire,su4_moire2}, where crossovers between SU($N$) models with different $N$ are also realized using external tunable fields \cite{delre_crossover,delre_crossover_altermagnets}, underscoring the multifaceted relevance of our results for state-of-the-art platforms for quantum technology, and opening opportunities to control and assess the correlation properties of quantum materials and their interplay with broken-symmetry phases. A reduction of correlations by increasing the number of components has been also revealed by the behaviour of the spin susceptibility of two-dimensional gases with long-range Coulomb interaction with valley degeneracy \cite{Marchi2009}.


To go beyond our results, it is essential to extend the analysis to account for nonlocal correlations in finite dimensions \cite{bellomia2024quasilocal,Held_2site_weakU} and multipartite entanglement, as witnessed by quench dynamics \cite{Costa&Hauke_multipartite_coldatoms} or measured by the mutual information between many natural orbitals \cite{Faraday_Natural_Orbitals}. 
While the former two avenues will surely strengthen the connection with experimental observations, the latter has the potential to deepen the understanding of the intricate connection between traditional quantum resources and the concept of fermionic nongaussianity \cite{Gottlieb2005,Gottlieb2007,Paris_nonGaussianity,GeometricEntropy_MIT_Vollhardt,Held_nonfreeness,LyuBu24_nonGaussianity,Sierant2025_FAF,Sierant2026_FAF,Matchgate_Review,Mele_NonfreenessBounds,Mele_NonfreenessLearning,Lami2025_nongauss_doping}.

\section*{Data Availability Statement}

All numerical data supporting the findings reported in this manuscript are openly available in \cite{data}.

\begin{acknowledgments}
We acknowledge financial support from the MUR via National Recovery
and Resilience Plan PNRR Projects No.\,CN00000013-ICSC and No.\,PE0000023-NQSTI, as well as via PRIN
2020 (Prot.\,2020JLZ52N-002) and PRIN 2022 (Prot.\,20228YCYY7) programmes. GB further acknowledges
support through the SFB Q-M\&S project of the FWF, DOI 10.55776/F86.
For open access purposes, the authors have applied a CC BY-NC SA public copyright license to any author accepted manuscript version arising from this submission.
The authors acknowledge the TU Wien Bibliothek for financial support through its open access funding programme.
The Flatiron Institute is a division of the Simons Foundation.
\vfill
\end{acknowledgments}

\bibliographystyle{quantum}
\bibliography{References}

\end{document}